\documentstyle[aps,prl,multicol,epsf]{revtex}
\begin{document}
\author{X. X. Yi$^{1,2}$\footnote{Electronic address:yixx@dlut.edu.cn},L. Zhou$^1$,
and H. S. Song$^1$}
\address{
$^1$  Department of Physics, Dalian University of Technology,
Dalian 116024, China\\
$^2$ Institute of Theoretical Physics, Chinese Academy of
Sciences, Beijing 100080, China }

\date{\today}

\title{ Entangled two cavity modes preparation via a two-photon process} \maketitle
\begin{abstract}
We propose a scheme for entangling two field modes in two high-Q
optical cavities. Making use of a virtual two-photon process, our
scheme achieves maximally entangled states without any real
transitions of atomic internal states, hence it is immune to the
atomic decay.
 \ \ \\
{\bf PACS number(s):03.67.Lx, 32.80.Wr, 42.50.-p}
\end{abstract}
\vspace{4mm}
\begin{multicols}{2}[]

Entanglement is one of the most characteristic features of quantum
systems and lies at the heart of the difference between the
quantum and classical multi-particle world. It is the phenomenon
that enables quantum information processing and computing
\cite{Ekert}. Beyond these and other related applications, complex
entangled states, such as the GHZ triplets of particles
\cite{Zeilinger} can be used for tests of quantum non-locality
\cite{Pan}. Moreover, the relaxation dynamics of larger entangled
states sheds light on the decoherence process and on the
quantum-classical boundary \cite{Zurek}. There are a lot of
proposal devoted to the preparation of quantum entangled states,
among them the idea for photon down-conversion process \cite{Pan},
with trapped ions \cite{Sackett}, for cavity quantum
electrodynamics \cite{Rauschenbeutel}, with macroscopic objects
\cite{Julsgaard} or for an optical fibre \cite{Silberhorn} has
been realized experimentally.

In the latter case, the entanglement results from the nonlinear
interaction between the two modes in an optical fibre. This is
closely connected to the recent advance of enhancing nonlinear
coupling via electromagnetically induced transparency(EIT)
mechanism \cite{Schmidt}. Measured value of the $\chi^{(3)}$
 parameter are up to six order of magnitude  larger than usual
 \cite{Hau}. This has  opened the door toward the application of
 this kind of nonlinear process to quantum information processing
   even for the very low photon-number
 case \cite{Harris}. In fact, there are several proposals for exploiting
 huge Kerr
non-linearities  to perform computation and quantum teleportation
\cite{Cochrane,Vitali} or for quantum non-demolition measurements
\cite{Sanders}. Apart from the Kerr nonlinearity, the experimental
achievement of atomic Bose-Einstein condensation(BEC) also provide
us  a chance to create many particle entanglement with nonlinear
interactions
\cite{Sorensen,Pu,Duan1,Duan2,Bigelow,Poulsen,Helmerson,Micheli}.
All these show that the non-linear interaction between different
quantum  modes is a valuable resource for quantum information
processing.

In this paper, we present a new theoretical scheme for entangling
two quantum modes in two high-Q optical cavities. Through a
virtual two-photon process,  an effective non-linear interaction
between the two modes can be established. By making use of the
virtual two-photon process, our new protocol significantly reduces
the effect of atomic spontaneous emission during the entanglement
preparation process.

Our system consists of two optical cavities as in Ref.
\cite{Plenio} and an atom system surrounded by the two optical
cavities. The axis of the two cavities are perpendicular  each to
other, the internal structure of the atom is depicted in Figure 1,
\vskip -1.2cm
\begin{figure}
\epsfxsize=7cm \centerline{\epsffile{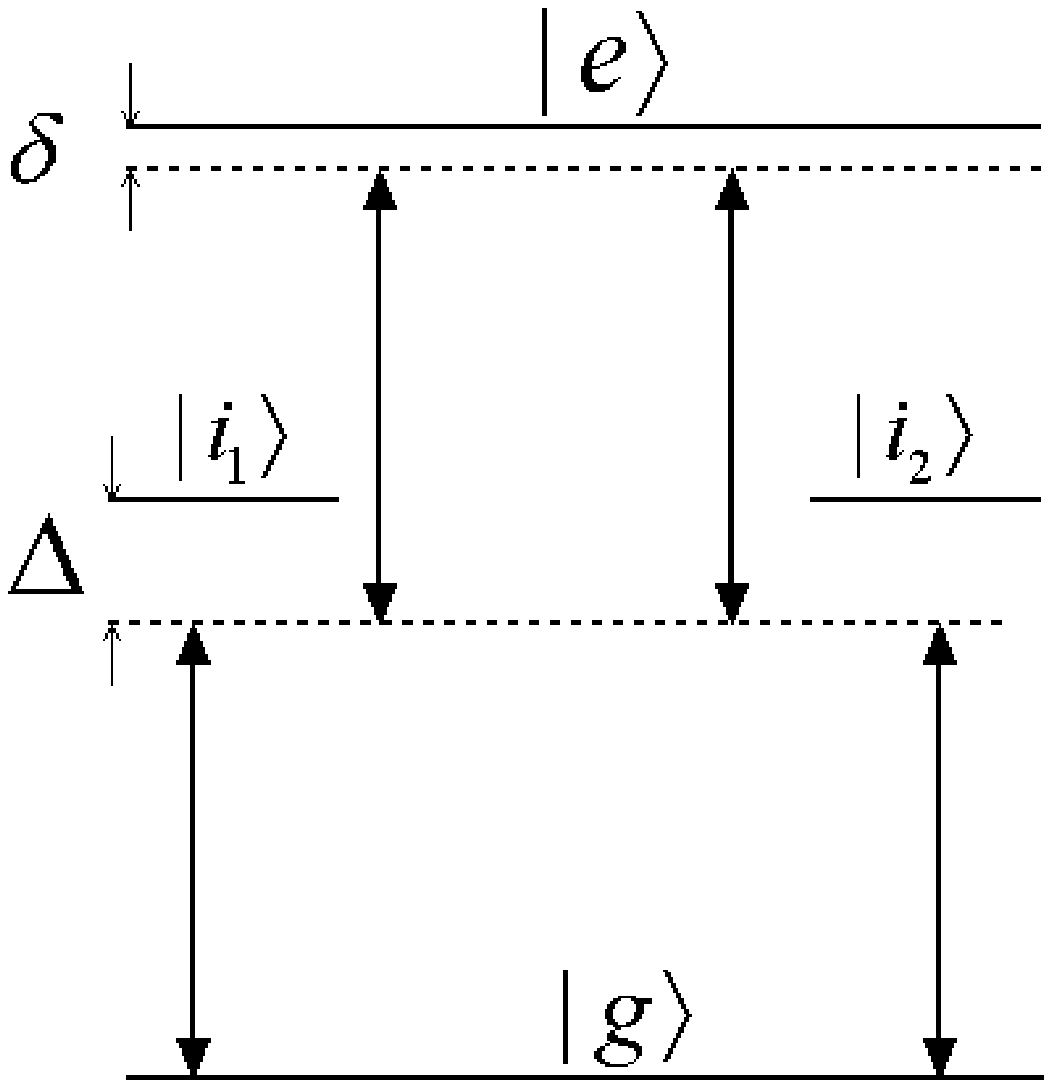 }} \vskip 1.5cm
\caption[]{A 4-level atom interacting with two cavity quantum
fields. Both cavity fields are detuned from atomic resonance by
$\Delta=\Omega-\omega_i$, and $\delta=2\Omega-\omega_e$.}
\end{figure}

 The atom is assumed to make
two-photon transitions of frequency $\omega_e$ between the
nondegenerate state $|g\rangle$ with energy $\omega_g=0$ and the
excited state $|e\rangle$. The transitions are mediated by two
intermediate degenerate levels $|i_1\rangle$ and $|i_2\rangle$
(with energy $\omega_i$): the frequencies for transitions
$|g\rangle\longleftrightarrow |i_1\rangle ( \mbox{or}
|i_2\rangle$) and $|i_1\rangle
(\mbox{or}|i_2\rangle)\longleftrightarrow |e\rangle$ are
$\Omega-\Delta$ and $\Omega+\Delta-\delta$, respectively. With
these notations, the system can be described by
\begin{eqnarray}
H&=&\hbar\Omega_a a^{\dagger}a+\hbar\Omega_b b^{\dagger}b
\nonumber\\
&+&\hbar g_c(|g\rangle\langle i_1|a^{\dagger}+|i_1\rangle\langle
e|a^{\dagger}+h.c.)\nonumber\\
&+&\hbar g_c(|g\rangle\langle i_2|b^{\dagger}+|i_2\rangle\langle
e|b^{\dagger}+h.c.)\nonumber\\
&+&\hbar\omega_i|i_1\rangle\langle
i_1|+\hbar\omega_e|e\rangle\langle e|+\hbar
\omega_i|i_2\rangle\langle i_2| \label{Ha1},
\end{eqnarray}
where $a(b)$ and $a^{\dagger}(b^{\dagger})$ are the annihilation
and creation operator for the cavity mode $a(b)$ with frequency
$\Omega_a$ $(\Omega_b)$, respectively, $g_c$ is the coupling
constant of the atom to the cavity mode $a(b)$ driving the
transition $|g\rangle\longleftrightarrow |i_1\rangle $ or
$|i_1\rangle \longleftrightarrow |e\rangle$ (
$|g\rangle\longleftrightarrow |i_2\rangle $ or $|i_2\rangle
\longleftrightarrow |e\rangle$ ). We will not consider the
position dependance of the cavity-atom coupling $g_c(\vec{r})$, a
good approximation in the Lamb-Dicke limit. For the reason of
simplicity, we assume $\Omega_a=\Omega_b=\Omega$ hereafter. Our
Scheme works in the following limit: 1)Both the cavity mode $a$
and $b$ are strong detuned, i.e., $\Delta=\Omega-\omega_i>>g_c,
\Delta>>\delta$ and $\delta >> |g_c|^2/\Delta$; and  2) the cavity
decay rete $\kappa<<|g_c|^2/\Delta$ as required for the high-Q
optical cavity. Because these transitions
$|g\rangle\longleftrightarrow |i_1\rangle $ , $|i_1\rangle
\longleftrightarrow |e\rangle$ , $|g\rangle\longleftrightarrow
|i_2\rangle $ and $|i_2\rangle \longleftrightarrow |e\rangle$
driven by the two cavity modes are far off-resonant, we may
adiabatically eliminate the intermediate states $|i_1\rangle$ and
$|i_2\rangle$ independently, the Hamiltonian (\ref{Ha1}) then
takes the following form \cite{Puri,Alexanian}
\begin{eqnarray}
\cal{H}&=&\hbar\omega a^{\dagger}a+\hbar\omega
b^{\dagger}b+\hbar\lambda(|g\rangle\langle e|a^{\dagger
2}+|e\rangle\langle g|a^2)\nonumber\\
&+&\frac{\omega_A}{2}(|e\rangle\langle e| -|g\rangle\langle
g|)\nonumber\\
 &+&\hbar \lambda(|g\rangle\langle e| b^{\dagger 2}+|e\rangle\langle
 g|b^2)\label{Ha2},
\end{eqnarray}
with $\omega=\Omega+2\frac{|g_c|^2}{\Delta}$,
$\Delta=\Omega-\omega_i$, $\lambda=\frac{|g_c|^2}{\Delta}$,
$\omega_A=\omega_e-\omega_g$. This is the Hamiltonian which is
broadly used to describe the two-photon process, and has received
an extensive study during the last decades. For instance, the
experimental realization of a two-photon cascade micromaser
\cite{Brune}, the generation of Squeezing amplification
\cite{Gerry} and the creation of entangled states
\cite{Cardimona}. Our proposal works with a new mechanism
different from that by making use of very high Kerr coupling, the
coupling between the two cavity modes in our protocol, to be
discussed below, results from virtual two-photon processes.

In the limit $\delta>>\lambda$,  i.e.,
$(2\omega-\omega_A)>>|g_c|^2/\Delta$, the two-photon process is on
off-resonance, we may adiabatically eliminate the atom from the
system, the Hamiltonian (\ref{Ha2}) then takes the following form
in the interaction picture
\begin{equation}
H_{eff}=\hbar\frac{|\lambda|^2}{\delta}(a^{\dagger
2}a^2+b^{\dagger 2}b^2+ a^{\dagger 2}b^2+b^{\dagger 2}a^2)
\label{Ha3},
\end{equation}
in derivation of the Hamiltonian (\ref{Ha3}), the atom in its
ground state $|g\rangle$ initially is assumed. The two mode states
will be defined in terms of the usual two-mode Fock states
$|m,n\rangle=|m\rangle_a\otimes |n\rangle_b$ with $m(n)$ photons
in mode $a(b)$. First we consider a simple case that there are
only two photons in the mode $a$ while the cavity mode $b$ in
vacuum initially. The Hamiltonian (\ref{Ha3}) for this simple case
is equivalent to
$$
H_{eff}=2\hbar\frac{|\lambda|^2}{\delta}(|E\rangle_a\langle
E|+|E\rangle_b\langle E|+\sigma_a^+\sigma_b^-+
\sigma_a^-\sigma_b^+),
$$
with definition  $|E\rangle_x=|2\rangle_x,\ \ (x=a,b)$, and
$\sigma_x^{+(-)}$ is the pauli operator for mode $x$. It shows
that after the  interaction time $T_0=\delta\pi/8|\lambda|^2$, the
two cavity modes evolve to a maximal entangled state
$\frac{1}{\sqrt{2}}(|0 E\rangle+|E 0\rangle)$ while leave the atom
in its ground state $|g\rangle$. We would like to note that the
non-linear interaction term $[a^{\dagger 2}b^2+h.c.]$ is different
from the Kerr nonlinear interaction $a^{\dagger 2}a^2$ and
$a^{\dagger}ab^{\dagger}b$ \cite{Agarwal,Scully}, as the latter
one is in the form of the square of the free Hamiltonian, and
hence  either $a^{\dagger}a$ or $b^{\dagger}b$ is a constant of
motion.

We have performed extensive numerical simulations with the full
Hamiltonian Eq. (\ref{Ha1}). Ignoring the atomic spontaneous
emission and the cavity decay, we find the above analytical
insights to be completely accurate, i.e., we indeed get the
maximal entangled state (see Figure 2). In fact, we find that the
approximated Hamiltonian Eq.(\ref{Ha3}) is quite an good approach
to the full Hamiltonian Eq.(\ref{Ha1}).

The top panel in Figure 2 shows selected results for the
dependance of the population   of state $|0,2\rangle$(dotted line)
and $|2,0\rangle$ (dashed line)  on time, while the lower panel
display the von Neumann entropy taken in base 2.  An initial state
$|0,2\rangle$ and   system parameters $\Delta=20g, \delta=5g$ are
chosen for this plot.  The maximal entangled state may be obtained
at time $t=785$ with perfect Fidelity ($>> 99.9\% $). The Fidelity
is defined as the overlap between the output state and the desired
state.

\begin{figure}
\epsfxsize=7cm \centerline{\epsffile{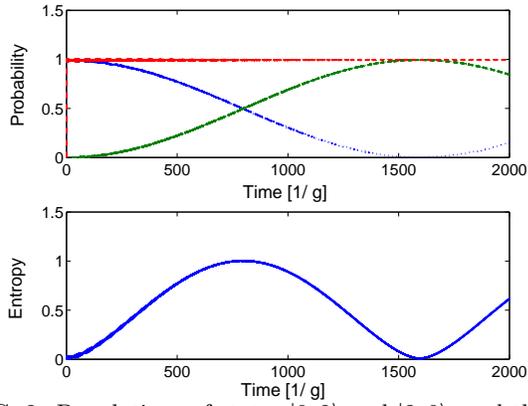}} \vskip 0cm
\caption[]{Populations of states $|0,2\rangle$ and $|2,0\rangle$,
and the von Neumann entropy versus time, this figure is plotted
for the case when the adiabatic limit satisfied. The constant line
in the top panel denotes the probability of the atom being in its
ground state, disregard where the photons are.}
\end{figure}
Similar results are found for the initial state $|4,0\rangle$, it
is illustrated in Figure 3. The difference is that there are three
components $|4,0\rangle$, $|2,2\rangle$ and $|0,4\rangle$ in the
output state, their population are shown in the top panel in
Figure 3 by c,b,a, respectively. It is interesting to note that
there is a time point(in Fig. 3, top panel) when line c and line a
overlap, which corresponds to the system in
$\frac{1}{\sqrt{2}}(|0,4\rangle+|4,0\rangle)$. And at this point
the entropy as plotted in the lower panel is 1, it equals the
entropy of a maximally entangled state for two-qubit. A recent
study on entangled two modes show that the amount of entanglement
present in a given state depends on how one defines one's
systems\cite{enk}. This means we could redefine our two modes such
that the state $\frac{1}{\sqrt{2}}(|0,4\rangle+|4,0\rangle)$
represents a maximally entangled state.
\begin{figure}
\epsfxsize=7cm \centerline{\epsffile{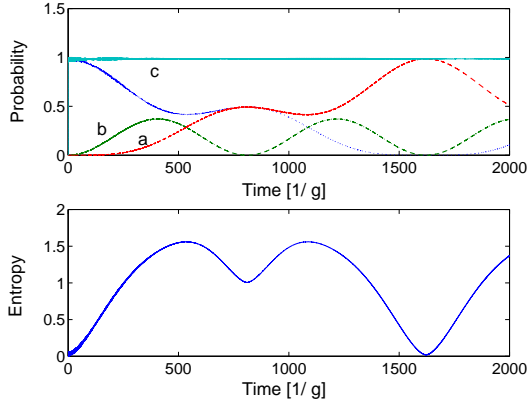}} \vskip 0cm
\caption[]{The same as in Fig.2, but for the initial state
$|4,0\rangle$}
\end{figure}

It is surprising to find that the same dynamics as in the
adiabatic limit persist even when adiabatic elimination is not
valid. As an example, in Figure 4 and Figure 5, we display results
for $\Delta=8g, \delta=3g$. Apparently, the atom as the
interaction agent  for the two cavity modes is enough for
establishing an effective interaction between them.

\begin{figure}
\epsfxsize=7cm \centerline{\epsffile{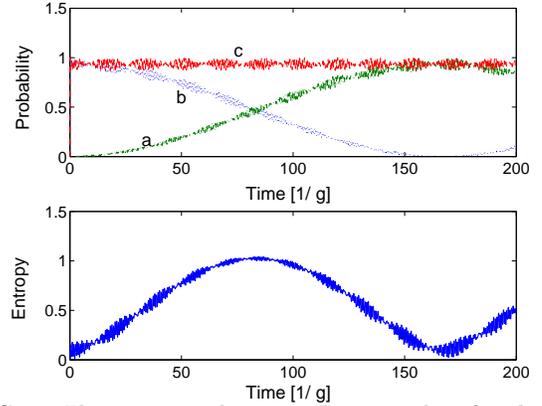}} \vskip 0cm
\caption[]{The same results as in Figure 2, but for the case where
adiabatic elimination of the atomic levels is not valid. }
\end{figure}
\begin{figure}
\epsfxsize=7cm \centerline{\epsffile{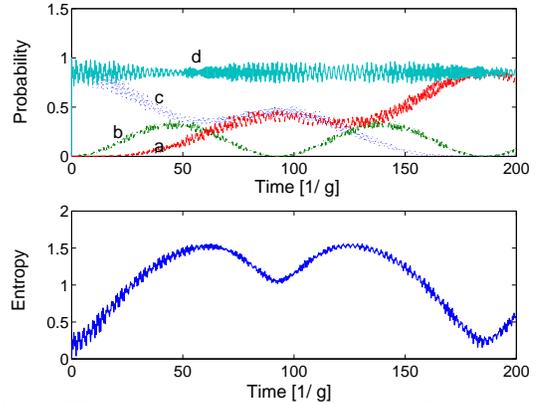}} \vskip 0cm
\caption[]{The same results as in Figure 3, but for the case where
adiabatic elimination of the atomic levels is not valid.}
\end{figure}

Now, we discuss effects of the dissipation or decoherence due to
both the atomic decay and the cavity loss. As with any proposal
for quantum information processing, ultimately its success depends
on being able to complete many coherent dynamics during the
decoherence time. In principle, as long as a)
$\frac{\lambda^2}{\delta}
>>\kappa$ and b) $\frac{\lambda^2}{\delta} >>\Gamma$, we could
expect essentially the same results as illustrated in Figure 2 and
Figure 3. As there are no real transitions of atomic states in our
proposal, it makes this scheme immune to the atomic spontaneous
emission or atomic decay, so the restriction b) make no sense in
this scheme. On the other hand, the condition a) is difficult to
achieved because the two-photon process is relatively weak due to
large off-resonant detunings for all its intermediate states. In
Figure 6 and Figure 7, the effects of cavity decay on dynamics of
the proposed system is illustrated. As known, the decoherence time
for a state $|m,n\rangle$ depends on the total number of photons,
and as Figure 6 and Figure 7 show, relative good results are found
when the  cavity loss rate $\kappa$  is small.
\begin{figure}
\epsfxsize=7cm \centerline{\epsffile{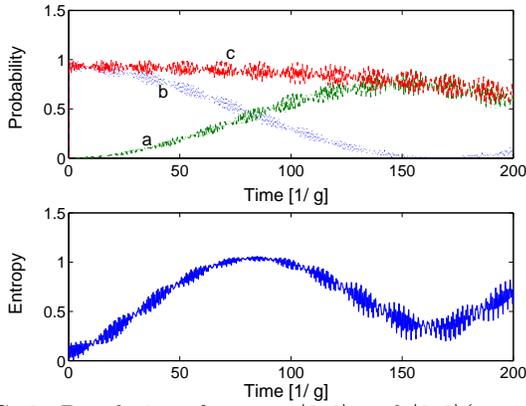}} \vskip 0cm
\caption[]{Populations for state $|2,0\rangle$ and
$|0,2\rangle$(top panel), as well as the von Neumann entropy, with
cavity decay rate $\kappa=0.005g$, the other parameters chosen are
the same as in Figure 4.}
\end{figure}
\begin{figure}
\epsfxsize=7cm \centerline{\epsffile{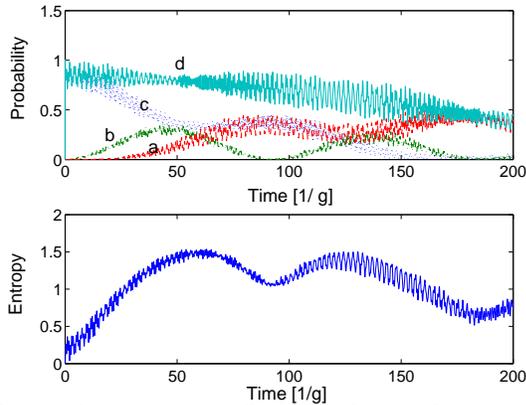}} \vskip 0cm
\caption[]{The same as in Figure 5, but with cavity decay rate
$\kappa=0.005g$. }
\end{figure}

Finally, we want to stress that the requirement for the
intermediate states degenerate is not necessary, in fact, our
proposal works in the same manner when the detuning $\Delta_i$
defined by $\Delta_i=\Omega-\omega_i$ have a different sign, i.e.,
$\Delta_1=-\Delta_2$. As all existing cavity QED-based quantum
computation protocols, optical cavity with high-Q and atom with
small decay rate remain challenging because of the technological
limit of the Fabry-Perot optical
cavity\cite{pellizzari,pachos,beige,you}.

In conclusion, we have proposed a new protocol for preparing the
maximally entangled state in two high-Q optical cavities. As the
photons act as the information carrier, cavity with very high-Q
factor is highly desired. We have explained the scheme in terms of
the virtual two-photon process induced nonlinear interaction. In
addition, our protocol can also be explained in terms of
entanglement distribution by separable states \cite{Cubitt}. This
new protocol has advantages that its successful implementation
involves no real transitions of atomic states, it makes this
proposal immune to the atomic spontaneous emission or atomic
decay.
\ \ \\
{\bf ACKNOWLEDGEMENT:}\\ This work is supported
by EYTP of M.O.E, and NSF of China.\\

\end{multicols}
\end{document}